\documentclass[letterpaper, 10 pt, conference]{ieeeconf}  % Comment this line out if you need a4paper

\IEEEoverridecommandlockouts                              % This command is only needed if 
\usepackage{graphics} % for pdf, bitmapped graphics files
\usepackage{epsfig} % for postscript graphics files
\usepackage{subfigure}
\usepackage{amsmath} % assumes amsmath package installed
\usepackage{amssymb}  % assumes amsmath package installed
\usepackage{subcaption}
\usepackage{multirow}
\newcommand{\RomanNumeralCaps}[1]
    {\MakeUppercase{\romannumeral #1}}

\title{\LARGE \bf
A Near-Real-Time Processing Ego Speech Filtering Pipeline Designed for Speech Interruption During Human-Robot Interaction}

\author{Yue Li$^{1}$, Florian A. Kunneman$^{2}$, and Koen V. Hindriks$^{1}$% <-this % stops a space
\thanks{*This work was supported by Chinese Scholarship Council}% <-this % stops a space
\thanks{$^{1}$Yue Li and Koen V. Hindriks are with Department of Computer Science, Vrije Universiteit Amsterdam, 1081 HV Amsterdam, The Netherlands.
        {\tt\small y6.li@vu.nl} {\tt\small k.v.hindriks@vu.nl}}%
\thanks{$^{2}$Florian A. Kunneman is with the Department of Languages, Literature and Communication, Utrecht University, 3584 CS Utrecht, The Netherlands.
        {\tt\small f.a.kunneman@uu.nl}}%
}

\begin{document}

\maketitle
\thispagestyle{empty}
\pagestyle{empty}

\begin{abstract}
%Limited by geometry design and the inability of 
%Finally, the limitations and future work of the proposed pipeline in real-time HRI are also discussed.
With current state-of-the-art automatic speech recognition (ASR) systems, it is not possible to transcribe overlapping speech audio streams separately. Consequently, when these ASR systems are used as part of a social robot like Pepper for interaction with a human, it is common practice to close the robot's microphone while it is talking itself. This prevents the human users to interrupt the robot, which limits speech-based human-robot interaction. To enable a more natural interaction which allows for such interruptions, we propose an audio processing pipeline for filtering out robot's ego speech using only a single-channel microphone. This pipeline takes advantage of the possibility to feed the robot ego speech signal, generated by a text-to-speech API, as training data into a machine learning model. The proposed pipeline combines a convolutional neural network and spectral subtraction to extract overlapping human speech from the audio recorded by the robot-embedded microphone. When evaluating on a held-out test set, we find that this pipeline outperforms our previous approach to this task, as well as state-of-the-art target speech extraction systems that were retrained on the same dataset. We have also integrated the proposed pipeline into a lightweight robot software development framework to make it available for broader use. As a step towards demonstrating the feasibility of deploying our pipeline, we use this framework to evaluate the effectiveness of the pipeline in a small lab-based feasibility pilot using the social robot Pepper. Our results show that when participants interrupt the robot, the pipeline can extract the participant's speech from one-second streaming audio buffers received by the robot-embedded single-channel microphone, hence in near-real time.
\end{abstract}

%%%%%%%%%%%%%%%%%%%%%%%%%%%%%%%%%%%%%%%%%%%%%%%%%%%%%%%%%%%%%%%%%%%%%%%%%%%%%%%%

%
%
%
\section{Introduction}\label{section:1}
% Problem description
During a human-like multi-party conversational interaction, the ability to handle overlapping speech signals, such as backchanneling and interruption, ensures efficient turn-taking between all parties \cite{c1}. For the same reason, in speech-based human-robot interaction (HRI) where a user and a robot try to have a natural conversation, the robot should listen %and process the received signal
all the time, implying that the microphone of the robot is always on, also when the robot is speaking \cite{c2}. In practice, however, this is difficult, as current social robots are not equipped to extract human speech while the robot is also talking. Current state-of-the-art automatic speech recognition (ASR) systems are unable to transcribe an overlapping speech audio stream and will instead most likely return a transcript of what the robot is saying itself. What is lacking is the ability of the robot to filter out its ego speech from the audio recorded by its microphone when it overlaps with the speech of a human user.

% Current state in HRI study
Therefore, it is common to adopt a simplex channel approach in HRI, which results in a rigid and unnatural turn-taking scheme in which the microphone is switched on and off repeatedly \cite{c4}. As an alternative to this type of interaction scheme, the adoption of a duplex channel has been explored to filter ego noise (the robot's own speech), for example, by using an additional headset that only picks up the speech from the target human user \cite{c3}, or additional microphones that may be placed either on or next to the social robot \cite{c5,c6} to enhance the target speech signal. In this work, we propose a pipeline that directly filters the ego speech from the signal recorded by the robot. The pipeline aims to extract the overlapping human speech directly from the signal received by the embedded robot microphone. Furthermore, to make this practical, it is important to develop a pipeline that can filter robot ego speech in \textit{real-time} from the recorded audio \cite{c2}. 

% Potential solution and current limitations
% Introduce the TSE.
% (problem: time for processing. focus on competitive speech estimation.)

\begin{figure}[ht]
    \centerline{
     \includegraphics[width=2.5in]{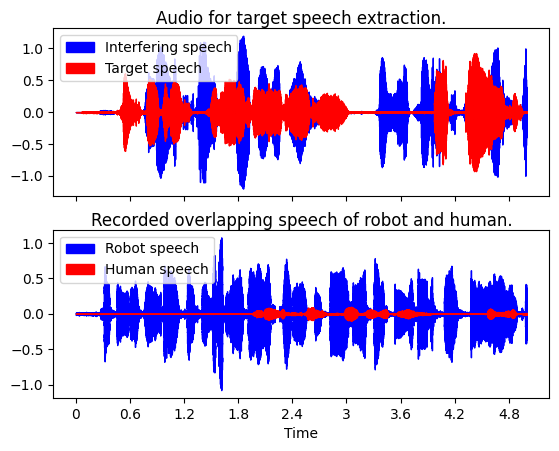}}
    \caption{Illustration of audio typically used for target speech extraction versus overlapping speech recordings of the Pepper robot (in time domain).}
    \label{fig:2}
\end{figure}

One idea to build such a pipeline is to use an existing \textit{Target Speech Extraction} (TSE) model. These models use %prior 
auxiliary information about the target speaker and extract their speech from the recorded mixture of different speakers and perhaps additional noise. The results reported in the literature are very promising \cite{c7}. There are, however, a few issues that prevent the use of these models for social robots in practice. First, current TSE models assume that the speech signal to be extracted has approximately the same or slightly less power than the interference speech \cite{c9}. However, this is often not the case for social robot platforms, where the distance between the microphone and the loudspeaker is small. Because of this, the robot's ego speech has typically much more power than the target speech in the received signal \cite{c3}, as illustrated in Fig.\ref{fig:2}. This severely hinders most TSE systems in extracting the target speech. Second, current TSE models expect a complete audio fragment to be processed. They have not been designed to incrementally extract the target speech from an ongoing audio stream from the robot's embedded microphone \cite{c7,c8,c9,c10}, which is essential for real-time human-robot interaction.

% Research Question
To address these issues, we propose a new pipeline in this paper that enables a robot to keep its microphone open and extract the interruption speech from a human user that overlaps with the robot ego speech in near-real time. Based on our previous work \cite{c3}, the basic idea of our pipeline is to take advantage of the fact that we know what speech signal the robot is to produce, instead of assuming that any characteristics about the target speech or the user interacting with the robot are known. Our aim is to address the following three research questions:
\begin{enumerate}
    \item \textit{Pipeline Design}: How can we design an ego speech filtering pipeline that filters out robot ego speech and preserves the overlapping human speech from the audio recorded by the robot's single-channel microphone? 
    \item \textit{Performance}:
    \begin{enumerate}
        \item \textit{Offline:} How does the proposed pipeline perform compared to two other state-of-the-art TSE models re-trained on the same overlapping human-robot speech dataset?
        \item \textit{Online:} How does the pipeline perform in a small lab-based feasibility pilot? Can it be used for real-time interaction? 
    \end{enumerate}

\end{enumerate}

% Contribution
Compared to our previous work \cite{c3}, the main contribution of this work is that we propose a convolutional neural network (CNN) that estimates the spectrogram of robot ego speech in the received audio, and combine it with spectral subtraction (SS) to obtain interrupted speech from an interacting human user. We evaluated the performance of this pipeline and compared it with our previous work and two other retrained models in the literature on the same dataset \cite{c3}. To assess the feasibility of using our pipeline in practice, we integrated it into the Social Interaction Cloud (SIC) software framework \cite{c26}\footnote{The Social Interaction Cloud (SIC) is a lightweight software framework developed with the aim of facilitating developers and researchers to (quickly) prototype a social robot application. SIC provides an interface for the Pepper robot used in this paper. Among other things, it provides several components to facilitate speech-based interaction such as Google TTS APIs, which enables the Pepper robot to use other voices than its built-in voice.}, which facilitates the easy deployment of our pipeline on a social robot and makes our work available to the broader community. We then used this framework to evaluate the effectiveness of our pipeline in a small lab-based feasibility pilot, where we also looked at whether our pipeline can be used in a near-real-time setting on the popular humanoid robot Pepper.

The remainder of this paper is organized as follows. In Section~\ref{section:2}, we discuss related work focusing on different approaches to TSE, which we use as inspiration for our pipeline. In Section~\ref{section:3}, we introduce the proposed pipeline to filter out robot ego speech. Section~\ref{section:4} discusses the experimental setup and metrics to evaluate the pipeline, while Section~\ref{section:5} presents and analyzes the results. Section~\ref{section:6} concludes and discusses several limitations of the proposed pipeline in real-life HRI, as well as future work.

\section{Related Work}\label{section:2}
\begin{figure}[t]
      \centering
      \includegraphics[width=2.5in]{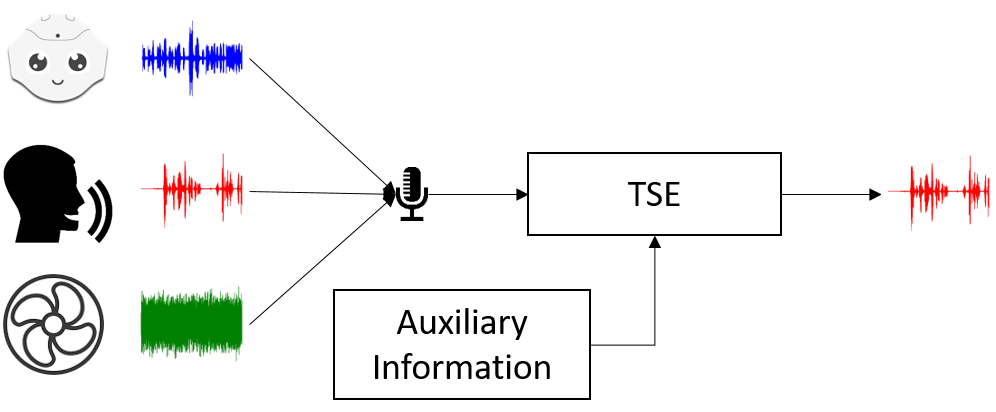}
      \caption{The problem of robot ego speech filtering}
      \label{fig:1}
\end{figure}

Fig.~\ref{fig:1} illustrates the main problem that a TSE system aims to address: extracting the target speech overlapping with robot ego speech from the audio recorded by the robot single-channel microphone. Here, we make a distinction between two main approaches: a signal processing-based (SP-based) approach and a deep learning-based approach.

\subsection{Signal Processing-Based TSE}
From an engineering point of view, the TSE problem in the single-channel approach is directly related to noise reduction and blind source separation \cite{c7}. SP-based TSE methods generally calculate and reduce audio noise in a spectrum space. The basic idea is that if noise $N_{ft}$ can be estimated, then an estimation of the target speech spectrogram $\hat{S}_{ft}$ can be obtained by subtracting the noise estimate from the sampled noisy spectrogram $X_{ft}$ \cite{c14}. A wide variety of approaches have been proposed to optimize the estimation of $\hat{S}_{ft}$, including spectral oversubtraction, multiband spectral subtraction, Wiener filtering, and iterative spectral subtraction \cite{c14}. These methods have been commonly designed and implemented to estimate $N_{ft}$ during pause or silence. They have shown their effectiveness in attenuating stationary noise \cite{c15}, and some have been tested to filter robotic fan noise or joint noise \cite{c16}. Other SP-based TSE methods, such as minimum mean-square estimation and the factorial hidden Markov model, are also investigated \cite{c13}. However, these methods require the estimation of the number of speakers and can lead to global permutation ambiguity \cite{c17}\footnote{An ambiguous permutation is a permutation which cannot be distinguished from its inverse permutation.}.

\subsection{Deep Learning-Based TSE}
The success in solving global permutation by deep-clustering \cite{c18} and permutation invariant training \cite{c19} leads to the emergence of proposed neural network-based TSE systems. Wang et al. proposed a speaker-open network VoiceFilter \cite{c8} and its following work, VoiceFilter-Lite \cite{c20} as plug-ins before ASR, which extracted a representation of the target speaker from a clean enrollment sequence and then isolated the speaker's voice in a mix. Similar ideas have been explored by Ge et al. \cite{c21}. Zmolikova et al. \cite{c9} later introduced SpeakerBeam, which explored three different methods of informing the network to modify the behavior of the acoustic model. To avoid the adverse effect on performance from different window lengths when analyzing the reference signal and the input mixture signal, Subakan et al. \cite{c10} and Luo et al. \cite{c22}, respectively, proposed two time-domain solutions for TSE, Sepformer and ConvTasNet. Although the performance of these proposed networks is promising in public datasets \cite{c23}, they have not been tested or evaluated in real recordings during HRI, where human speech overlaps with robot speech in HRI.

In our previous work \cite{c3}, we proposed two ego speech filtering methods, SP-based and neural network-based (NN-based). The SP-based method calculated the loudspeaker-microphone response function to map the speech spectrogram generated by the text-to-speech (TTS) API to the spectrogram of the microphone-picked signal and estimated the target speech by subtracting it from the spectrogram of the overlapping speech. The NN-based method consisted of a convolutional recurrent network that directly adopted the spectrograms of the mixture audio and generated speech audio as input. The result was not satisfactory. In the speech spectrogram extracted by the SP-based method, there were residues of robot speech caused by reverberation. The NN-based method was proved to be robust to reverberation, but there were frames not predicted. Nevertheless, the improved WER showed that the design strategy was feasible, which mapped the spectrogram of the robot-generated speech to that of the robot-recorded speech and subtracted it from the spectrogram of the recorded overlapping speech. In this work, we followed the same strategy and aimed to improve performance.

\section{Method}\label{section:3}
\subsection{Problem Formulation}
\label{section:3.1}
%
% The target application scenario is illustrated in Fig.~\ref{fig:3}. 
% \begin{figure}[b]
%       \centering
%       \includegraphics[width=2.0in]{pics/physical_environment_demonstration.png}
%       \caption{Setting that is the focus of this paper}
%       \label{fig:3}
% \end{figure}
The spectrum of the received single-channel audio $X_{ft}$ can be modelled by  Equation~\ref{eq:1}:
\begin{equation}
\label{eq:1}
    X(f,t) = S(f,t) + N(f,t)
\end{equation}
where $S_{ft}$ is the short-time Fourier transformed (STFT) spectrum of the target speech signal, that, in our case, consists of speech that is interrupting or overlapping with robot speech, and $N_{ft}$ is the noise signal that in our case is composed of the overlapped robot speech, reverberation of the robot speech, and fan noise; $f$ and $t$ respectively denote the frequency and time indices. We denote the spectrogram of the speech signal to be played by the robot as $R_{ft}$. In this paper, our objective is to extract the target speech $S_{ft}$ from the recorded audio $X_{ft}$ with the generated robot speech $R_{ft}$. However, $R_{ft}$ cannot be directly subtracted from $X_{ft}$ due to the inconsistent and non-linear response of the speaker and microphone at different frequency levels \cite{c3}. Furthermore, the reverberation of the room will cause the value $R_{f_0t_0}$ to appear again at time $t_1$ ($t_1>t_0$) \cite{c24}. Therefore, a function $G(\cdot)$ must be learned to map $R_{ft}$ to an estimate of the recorded ego speech spectrogram $\hat{R}_{ft}$. The estimate $\hat{S}(ft)$ of the spectrogram of the overlapping human speech can then be obtained by Eq.~\ref{eq:2}:
\begin{equation}
    \hat{S}(f,t) = X(f,t) - G(R(f,t))
\label{eq:2}
\end{equation}
\subsection{Proposed CNN}

\begin{figure}[t]
    \centering
    \subfigure[The overall architecture.]{\includegraphics[width=0.30\textwidth]{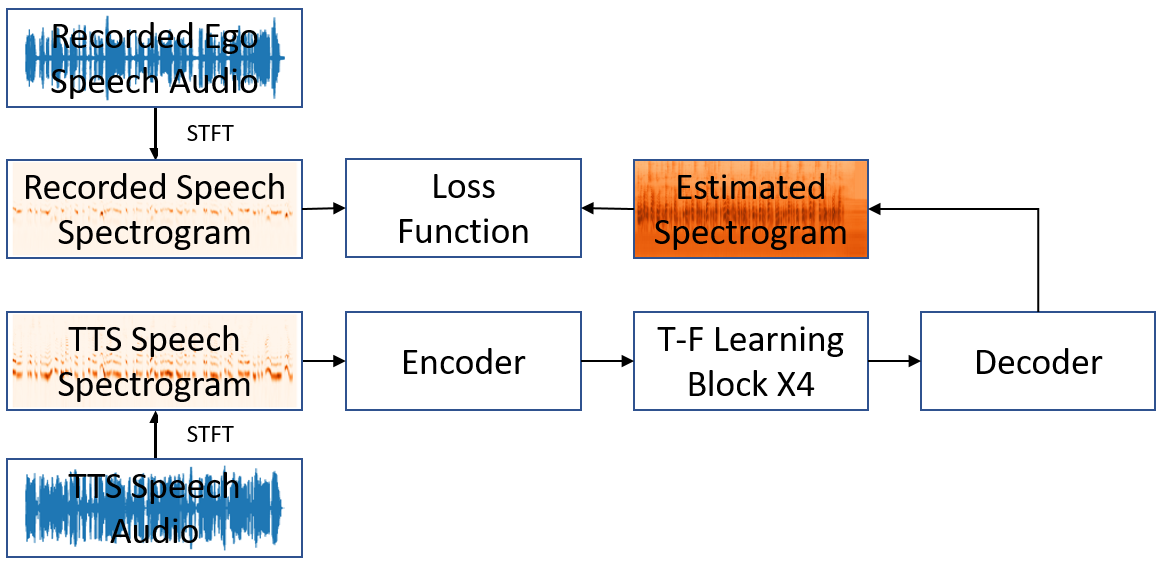}}
    \hfill
    \subfigure[The T-F learning block.]{\includegraphics[width=0.40\textwidth]{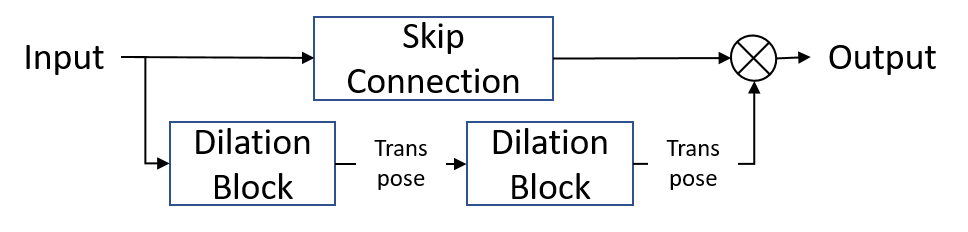}}
    \caption{The architecture of the proposed CNN.}
    \label{fig:5}
\end{figure}

In this work, we propose a CNN to learn $G(\cdot)$, to estimate the spectrogram of robot ego speech, as shown in Fig.~\ref{fig:5}. 
This network predicts the magnitude spectrogram $\hat{R}_{ft}$ from the magnitude spectrogram $R_{ft}$. It is trained to minimize the difference between the prediction and the magnitude spectrogram of the recorded robot ego speech. The network architecture consists of three blocks: an \textit{encoder}, a \textit{T-F Learning} block, and a \textit{decoder}. The encoder block is a $5\times5$ 2-dimensional (2D) convolutional layer with 128 channels and $2\times2$ padding, attached with a ReLU activation layer. The decoder block is a transposed $5\times5$ 2D convolutional layer with 128 channels and $2\times2$ padding, attached to a sigmoid activation layer. 
\begin{figure*}[t]
      \centering
      \includegraphics[width=5.0in]{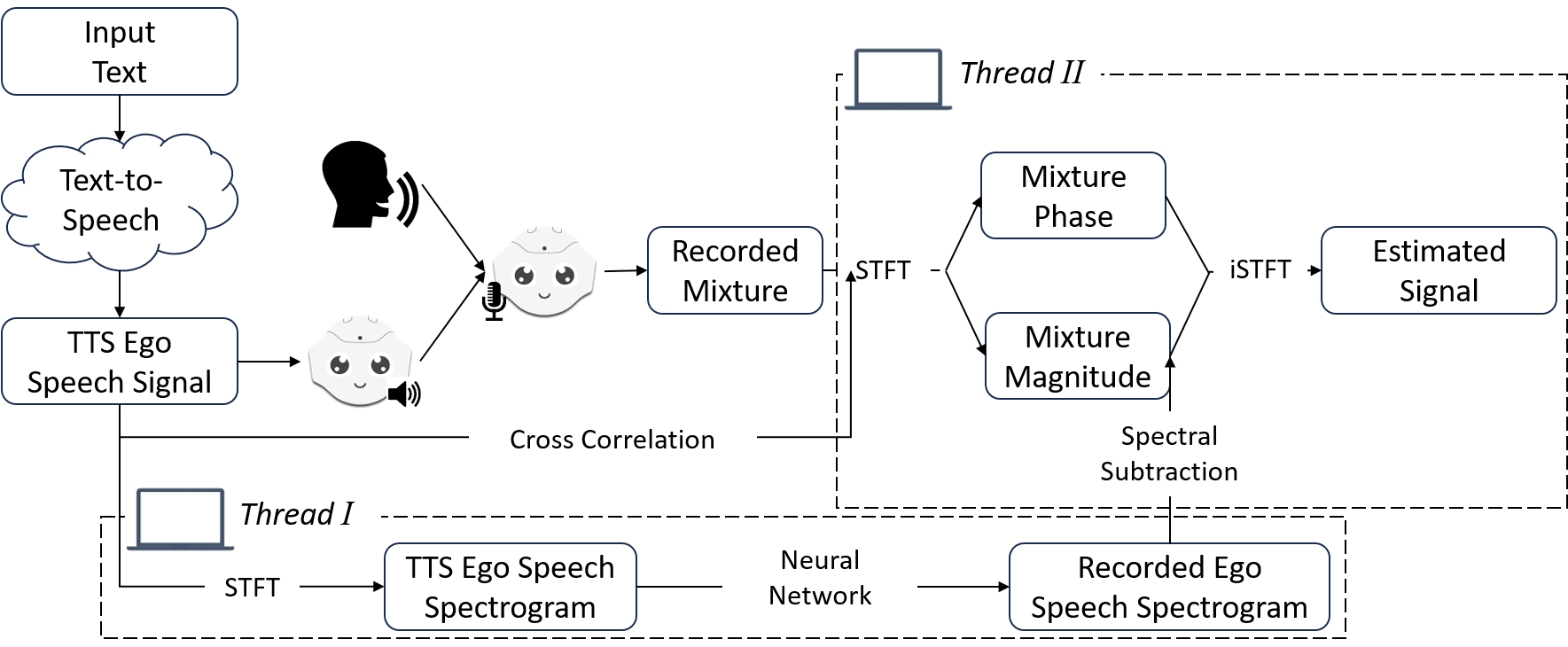}
      \caption{The proposed robot ego filtering pipeline}
\label{fig:4}
\end{figure*}
The T-F Learning block is designed to learn relations between the frequency axis and the time axis, respectively. It has two \textit{Dilation} blocks and one \textit{Skip Connection} block. The \textit{Skip Connection} block is designed to ensure feature reusability from the last layer \cite{c25}. It is a $1\times1$ 2D convolutional layer with 128 channels followed by a \textit{ReLU} activation layer. Each \textit{Dilation} block is designed to learn the information between the frequency axis and the time axis separately. Each has one $5\times5$ 2D convolutional layer with 128 channels and an incremental dilation along the last dimension, followed by \textit{ReLU} activation layer. The input is padded %before fed in 
to make the output size the same as the input. After each \textit{Dilation} block, the output is transposed between the last two axes. %This design is to ensure that each block learns the relations between time frames and frequency bins, respectively. 
The dilation hypermeter is set to $[2,4,8,16]$ for each block to increase the receptive field of each kernel. The number of parameters in the network is 1.71MB.

% During training, all input audio fragments are truncated to a 5-second length and converted to a single channel with a sampling rate of 16k~Hz. 

%
%

\subsection{Pipeline Design for Real-time Processing}
In the SIC framework, once the command is sent to the robot to make it speak, the speech content is sent to either Google \cite{c27} or its embedded TTS API. Afterwards, the entire corresponding speech audio is generated, sent to the local computer, and then uploaded to the robot. The robot will then play this audio through its loudspeaker, and the microphone will remain open and recording throughout the speech. The proposed real-time processing robot ego speech filtering pipeline is illustrated in Fig.~\ref{fig:4}.

We create two threads in the pipeline. In \textit{Threat} \RomanNumeralCaps{1}, we preload the proposed and trained CNN and perform STFT on the generated speech audio. Its magnitude spectrogram $R_{ft}$ is sent to the network. The trained CNN then predicts the corresponding estimated magnitude spectrogram $\hat{R}_{ft}$ and we send it to \textit{Thread} \RomanNumeralCaps{2}.

In \textit{Thread} \RomanNumeralCaps{2}, we create an audio buffer of one second to store the recorded audio stream for further processing.% As discussed in our previous work, the alignment of the time between $X_{ft}$ and $\hat{R}_{ft}$ is crucial for the performance of SS in TSE. Nevertheless, the latency of the Internet causes the time difference to be inconstant between the time the command is sent and the time the robot starts playing the audio. 
We adopt the same voice activity detection method proposed in our previous work \cite{c3} to align the time between $X_{ft}$ and $\hat{R}_{ft}$, empty the streamed data in the buffer before speech starts, and pad with new streaming data. We then perform STFT on this audio buffer and get the magnitude spectrogram $X_{ft}$ and the phase spectrogram $\phi_{ft}$. After subtracting the corresponding frames in $\hat{R}_{ft}$ from $X_{ft}$, the estimated speech signal can be obtained by performing inverse STFT (iSTFT) on the combination of the estimated magnitude spectrogram $\hat{S}_{ft}$ and the unprocessed $\phi_{ft}$. To avoid the truncation effect caused by windowing, only the middle 0.8 seconds of the estimated speech signal are kept and divided by its maximum value \cite{c31}. We then store it or stream it to the ASR system for further analysis. To avoid repetition of calculations or omission of information, we empty the first 0.8 seconds of the data in the buffer, pad it with the new streaming data, and restart the estimation again until the end of the speech.

\section{Experiments}\label{section:4}
We evaluate the proposed pipeline offline by means of a test dataset and online in a small lab-based pilot.

\subsection{Dataset Generation and Offline Evaluation}
As discussed in Section~\ref{section:1}, existing public overlapping speech datasets are not valid for this study, where target speech possess the similar power as interfering speech. So, for offline evaluation, we adopted the same dataset from our previous work \cite{c3}, which used the public datasets Robot Voice\footnote{https://osf.io/v4y6h/} and Librispeech\footnote{https://www.openslr.org/12} \cite{c28}. %Robot Voice dataset contains 7913 single-channel real recorded ego speech of robot Pepper at a considerably loud volume in two rooms with different reverberation conditions, a large lab with weak reverberation and a small office with relatively strong reverberation. It also contains the corresponding generated speech audio of 17 different voice identifications, including Pepper's embedded voice and 16 generative human voices. The speech content was chosen from Librispeech randomly. 
%We adopt the same generation strategy as \cite{c3} because the overlapping speech generated is close to the real recordings by the embedded microphone of Pepper. 
We evaluate our proposed method in two different ways. First, we used the entire duration of the mixture as input for TSE, the same as that of a typical TSE system. Second, in order to evaluate the proposed near-real-time processing pipeline, we cut the mixture into 1 second long blocks with a 200-millisecond stride, to simulate the input as the streaming audio buffer. We then performed TSE on each block. After being divided by its maximum value, the output of each block was concatenated as the human speech extracted. The window length for STFT and iSTFT is set at 25~ms with a 10~ms stride.

\subsection{Network Training}
Based on the distribution in room size, 200 segments recorded in the large lab and 600 in the small office are randomly selected as the test set to evaluate different approaches, while the other segments are selected for the training set. We adopt the power law compressed reconstruction error \cite{c8} as the loss function to train our proposed network and VoiceFilter, the scale-invariant signal-to-distortion ratio \cite{c29} as the loss function to train ConvTasNet, and monitored all training processes with \textit{Tensorboard} to avoid overfitting. We set the learning rate at 0.001 and adopt Adam as the optimizer.

\subsection{Baseline}
As evaluated in \cite{c3}, the pre-trained TSE system, VoiceFilter, could not filter out the robot's speech from the recorded mixture. To fairly compare the performance of the proposed network with other state-of-the-art TSE systems, we retrain VoiceFilter, provided by Seuong-won Park\footnote{https://github.com/maum-ai/voicefilter} with 18.88MB parameters, and ConvTasNet, provided by Kaituo Xu\footnote{https://github.com/kaituoxu/Conv-TasNet/tree/master} with 4.47MB parameters, on the same training set and evaluate them on the same test set. To train and evaluate VoiceFilter, we use another recorded segment from Robot Voice as auxiliary information, which shares the same voice identification as the robot speech in the mixture. For ConVasNet, which originally predicts multiple estimations from different speakers, we change the last layer to output only the overlapping target human speech. We also report on the performance of public VoiceFilter without retraining, as well as the NN-based and SP-based methods in our previous work for comparison.

\subsection{Small Lab-Based Feasibility Pilot}
To verify the effectiveness of the proposed pipeline in the real-life HRI scenario, we integrate it with the SIC framework and conduct a pilot study of a laboratory-conditioned HRI experiment based on robot Pepper. The aim is to verify whether this pipeline can filter out Pepper's ego speech from its recordings and how well it can interpret the human speech contents.

\subsubsection{Participants}
We recruited 11 participants (6 males, 5 females) for our experiment. They are all randomly chosen undergraduates and graduates from the school campus who can fluently speak English. Consent forms for permission to record audio and video for analysis and demonstration are signed prior to the experiment.

\subsubsection{Procedure}
Before the experiment started, we informed the participants that Pepper was designed to %would play the role as an assistant in a shopping center 
and help people find the correct direction to a store in a shopping center. We asked the participants to stand or sit in front of Pepper at a distance where they felt comfortable interacting with it and instructed them to ask for directions to a place they want to go. We also informed the participants that Pepper would misinterpret their intentions and give the wrong direction to a wrong store (in this setup, we chose Adidas or Etos). Participants were asked to use speech to stop Pepper during its speech. We recorded the entire interaction and collected the estimated audio for further analysis. After the interaction, we asked the participants to watch the recorded video and help transcribe the words they used during the interruption. We then asked them to repeat these words as closely as they said during the interruption and recorded the clean speech using one Pepper-embedded microphone without overlapping robot speech. The generative voice identity was "en-US-Neural2-G", a female voice from the Google Text-to-Speech API. The entire incorrect instruction took approximately 8 seconds.

% \subsubsection{Logs and Transcripts}

%
\subsection{Evaluation Metrics}
To evaluate the performance in interpreting the content of overlapping human speech between the proposed pipeline and the retrained baseline networks, we use the word error rate (WER) as the evaluation metric for the offline evaluation. For calculation, we adopt Whisper\footnote{https://huggingface.co/openai/whisper-large-v3} as the ASR system to translate all the estimated and the original human speech and use the translation of original human speech as the ground truth. We report on the mean, median, and standard deviation (SD) values of the WER. We also report on the percentage of files whose WER is lower than 10\% and 50\%. 

For the small lab-based feasibility pilot, we use the WER and the computing time for each thread as the evaluation metric. The WER is to evaluate the ability of the proposed pipeline to restore human speech content. The ASR result of clean speech will be taken as the ground truth. In addition, computing time is used to evaluate the ability of the proposed method to process in real time. 

All calculations for evaluation are performed on a local desktop with an Intel(R) Core(TM) i9-9900K CPU and a NVIDIA GeForce RTX 2070 SUPER GPU.

\section{Results and Analysis}
\label{section:5}

\subsection{Offline Evaluation}

In Table~\ref{tab:1}, we present the results of the proposed method compared to the baseline models, with the original unfiltered overlapping speech data as a reference. The bold fonts in the table represent the best result.
\begin{table*}[ht]
\caption{Offline evaluation results}
\label{tab:1}
\begin{center}
\begin{tabular}{lccccccc}
\hline
\multirow{2}{*}{Method} & \multirow{2}{*}{Input} & \multirow{2}{*}{Retrain} & \multicolumn{3}{c}{WER /\%} & \multicolumn{2}{c}{Percentage of files /\%} \\ \cline{4-8} 
 &  &  & \multicolumn{1}{c}{Mean} & \multicolumn{1}{c}{Median} & SD & \multicolumn{1}{c}{\textless{}10\%} & \textless{}50\% \\ \hline
Unfilterd & - &X & \multicolumn{1}{c}{138.01} & \multicolumn{1}{c}{120.63} & 84.91 & \multicolumn{1}{c}{0} & 0 \\ 
SS-based & Blocks& X & \multicolumn{1}{c}{57.82} & \multicolumn{1}{c}{60.00} & 82.58 & \multicolumn{1}{c}{8.61} & 40.00 \\ 
NN-based & Entire& X& \multicolumn{1}{c}{67.90} & \multicolumn{1}{c}{75.38} & 31.60 & \multicolumn{1}{c}{6.07} & 29.02 \\ 
ConvTasNet& Entire & \checkmark & \multicolumn{1}{c}{23.53} & \multicolumn{1}{c}{17.86} & 21.57 & \multicolumn{1}{c}{32.07} & 88.49 \\ 
VoiceFilter & Entire & X & \multicolumn{1}{c}{131.9} & \multicolumn{1}{c}{113.8} & 82.58 & \multicolumn{1}{c}{0} & 0 \\ 
VoiceFilter & Entire & \checkmark & \multicolumn{1}{c}{19.53} & \multicolumn{1}{c}{15.38} & 18.54 & \multicolumn{1}{c}{37.34} & 93.09 \\ 
Proposed & Entire & \checkmark & \multicolumn{1}{c}{\textbf{14.43}} & \multicolumn{1}{c}{\textbf{7.69}} & \textbf{17.82} & \multicolumn{1}{c}{\textbf{55.56}} &\textbf{94.78} \\ 
Proposed & Blocks & \checkmark & \multicolumn{1}{c}{17.41} & \multicolumn{1}{c}{10.53} & 19.76 & \multicolumn{1}{c}{48.82} & 92.26 \\ \hline
\end{tabular}
\end{center}
\end{table*}

We can observe that the VoiceFilter fails to extract the target speech from the mixture without retraining. After retraining, its performance on the test dataset is relatively stable with an average WER value of 19.53\% (SD=18.54\%). Nevertheless, only 37.34\% of its predictions result in less than 10\% WER. The retrained ConvTasNet has a mean WER of 23. 53\% (SD=21.57\%). However, only approximately 32\% of its predictions can achieve a WER less than 10\%. In comparison, when TSE is performed on the entire mixture, our proposed method achieves the best result on all evaluation standards. Its extracted speech has less than 15\% mean WER (SD=17.82\%). In more than half of its predictions, only 10\% of the words are incorrect. And in almost 95\% of its predictions, less than half of the words are incorrectly translated. Furthermore, when human speech is extracted from the blocks% of mixture
, our proposed pipeline also obtains the second-best average and median WER values. In almost $50\%$ of its predictions, the ASR interprets less than 10\% words incorrectly, and 92.26\% of the predictions have more than half of the words correctly translated.
\begin{figure}[tbp]
    \centering
    \includegraphics[width=2.5in]{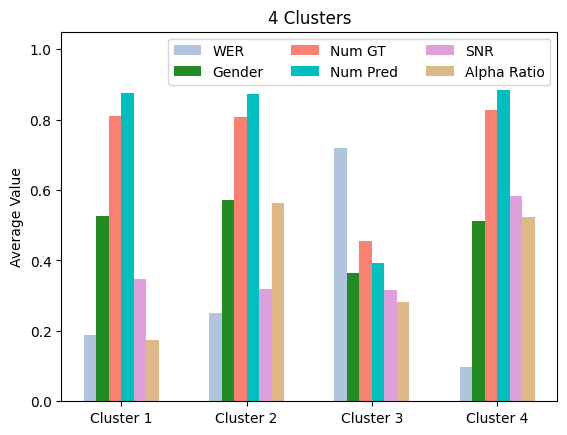}
    \caption{Results of agglomerative clustering.}
    \label{fig:6}
\end{figure}

\begin{figure}[tbp]
    \centering
    \subfigure[File index: 3612.]{\includegraphics[width=0.4\textwidth]{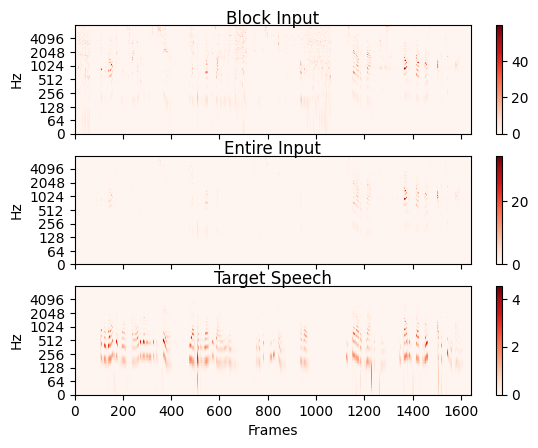}\label{fig:7a}} 
    \hfill
    \subfigure[File index: 6070.]{\includegraphics[width=0.4\textwidth]{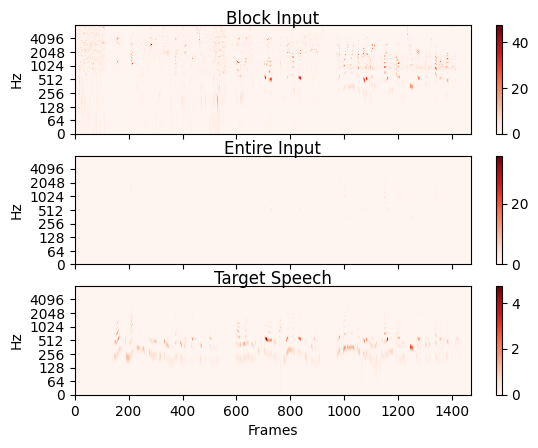}\label{fig:7b}} 
    \hfill
    \caption{Comparison of the estimated speech spectrograms.}
    \label{fig:7}
\end{figure}

%The results of the further analysis are listed below.
To inquire into challenging conditions for TSE, we analyze what characteristics the processed sound fragments are most related to a poor WER performance by applying Agglomerative Hierarchical clustering. The following features were included in this analysis: the number of words in the target speech, the number of words in the estimated speech, the SNR of the target speech during overlapping, and the alpha ratio (AR) of the target speech, which we define as Eq.~\ref{eq:3}. 
\begin{equation}
    AR = \frac{\sum(\sum_{i=1001}^{fs/2}(20\times log_{10}||S(f_i,t_j)||))}
    {\sum(\sum_{i=0}^{1000}(20\times log_{10}||S(f_i,t_j)||))}
  \label{eq:3}
\end{equation}
where $fs$ is the sampling rate, which is 16k~Hz. The higher the value of AR is, the more power the speech possess in high frequency bins. %Moreover, we select and perform STFT on two groups of estimated human speech. In the first group, the WER of the speech estimated by our proposed method on the entire mixture is significantly larger than that estimated by our proposed method on the input of the blocks; conversely, the trend is reversed in the second group. We compare their magnitude spectrograms for more information.
We found that the point in the dendrogram with four clusters gives the most insightful patterns, which are presented in Fig.~\ref{fig:6}. WER is included as information, with blocks of mixture. We also add information on the percentage of female (0) and male (1) speakers in each cluster. The average SNR of the target speech that overlaps with the robot speech is -22.33~db (SD=4.09). In comparison, classic TSE systems focus on extracting the target speech from the overlapping audio, where the SNR of the target speech is no less than -5~dB. 
%and the mean AR is 3.47\% (SD=0.10\%). The mean number of ground-truth words is 63.36 (SD=1.78), and the number of words in the extracted speech is 65.02 (SD=3.02). %Because we care more about the performance of the proposed pipeline in real-time processing, we include the WER of the extracted speech from the proposed method with blocks of mixture as input and the gender of the target speaker as references in Fig.~\ref{fig:6}, where we label female as 0, male as 1. However, we did not include them as features during the clustering calculation.

A comparison of spectrograms is presented in Fig.~\ref{fig:7}. The WER of the extracted speech in each selected file is as follows. File index 3612: 8.82\% for the blocks as input, 58.82\% for the entire mixture as input; File index 6070: 97.73\% for the blocks as input, 11.36\% for the entire mixture as input. 

We will offer further interpretations of Fig.~\ref{fig:6} and ~\ref{fig:7} in Section~\ref{section:5c}.

\subsection{Small Lab-based Feasibility Pilot}
In Table~\ref{tab:2}, we present the performance of the proposed pipeline during the laboratory-conditioned HRI. The translation mean WER of the extracted target speech is 8.39\% (SD=8.25\%). Regarding the computing time in each thread, we find that the processing time in \textit{Thread} \RomanNumeralCaps{1} is 980 ms, less than 1 second, while the computing time is as low as 10 ms in \textit{Thread} \RomanNumeralCaps{2}.
\begin{table}
\caption{Performance lab-based pilot}
\label{tab:2}
\begin{center}
\begin{tabular}{ccccc}
\hline
\multicolumn{3}{c}{WER}  & \multicolumn{2}{c}{Computing Time /ms}  \\ \hline
\multicolumn{1}{c}{Mean} & \multicolumn{1}{l}{Median} & SD  & \multicolumn{1}{c}{\textit{Thread} \RomanNumeralCaps{1}} & \textit{Thread} \RomanNumeralCaps{2} \\ 
\multicolumn{1}{c}{8.39\%} & \multicolumn{1}{c}{12.50\%}  & 8.25\% & \multicolumn{1}{c}{980}      & 10      \\ \hline
\end{tabular}
\end{center}
\end{table}

Among the 11 participants, the target speech of 5 male participants was perfectly restored and the WER of their speech is 0\%, and the WER of the other speech was also less than 20\%. %Table~\ref{tab:3} presents detailed information on misinterpreted speech. 
In each misinterpreted human speech, there are no more than 2 incorrectly predicted words. 4 out of 8 mistranslated words are the names of a brand, which themselves are challenges for the ASR system. 
%while 1 is a location. 1 backchanneling word is missing in the prediction, while the last comprehends the intention of the human oppositely by mistranslating "was talking" to "wasn't talking".
\begin{table}
\caption{Detailed information about the ASR result.}
\label{tab:3}
\begin{center}
\begin{tabular}{ccccc}
\hline
\multirow{2}{*}{Num} & \multirow{2}{*}{Gender} & \multirow{2}{*}{WER /\%} & \multicolumn{2}{c}{Incorrectly Predicted Words}          \\ \cline{4-5} 
 &   &       & \multicolumn{1}{c}{Ground Truth}  & Predicted   \\ \hline
1     & F & 12.5  & \multicolumn{1}{c}{"Nike"}        & "90"        \\ 
3 & F & 18.18 & \multicolumn{1}{c}{"Hema","Hema"} & "Ema","Ema" \\ 
6 & F & 14.29 & \multicolumn{1}{c}{"Adidas"}      & "anidots"   \\ 
8 & F & 18.18 & \multicolumn{1}{c}{"bookstore"}   & "box now"   \\ 
10 & M & 16.67 & \multicolumn{1}{c}{"was talking you"} & "wasn't talking" \\ 
12    & F & 12.5  & \multicolumn{1}{c}{"Eh"}          & ""          \\ \hline
\end{tabular}
\end{center}
\end{table}

\subsection{Analysis and Discussion}

\subsubsection{Offline Evaluation}
\label{section:5c}
% Compared to our previous work and the baselines,  the performance of the pipeline proposed in this work, which performs TSE directly on the entire mixture or on the blocks of mixture, is better and more reliable.
From Table~\ref{tab:1}, we can find that the TSE system with a raw audio wave as input performs worse than the systems with a spectrogram as input. This is because the first system extracts the target speech based on wave information, which can be hidden by robot speech in the recorded mixture, as shown in Fig.~\ref{fig:1}. We also find that the proposed system based on the generated speech signal as auxiliary information performs better than that based on speaker identification. This is because the performance of the speaker identification-based system depends on the values in the spectrogram. Nevertheless, compared to the robot interference speech, some values in the spectrogram of the target speech are too small during normalization and may be discarded from prediction. It will cause some of the frames to be incorrectly predicted in the extracted speech and further results in recognition failure.

From Fig.~\ref{fig:6}, we can observe that the SNR of the target speech in the mixture is the decisive factor for WER. Furthermore, when there are fewer words in the target speech, the WER will increase significantly. We also find that the WER is not obviously related to the speaker's gender. However, if a female speaks fewer words and her speech does not have a high AR value, it will be poorly translated, even at the same volume. On the contrary, the speech will be better translated when it is long and has a higher volume and AR value. This finding aligns with the results in \cite{c30}, where the authors found that the ASR system would translate female speech relatively worse than male speech in a noisy environment.

When comparing the speech spectrograms extracted from the proposed method with different inputs in Fig. \ref{fig:7}, we find that the values between 0 and 500~Hz are rarely restored. This is because the values in this range are shared by all speech, and spectral subtraction can easily cause oversubtraction. However, as discussed in \cite{c3}, the values in high-frequency bins are essential for speech recognition. The proposed pipeline can preserve and restore these values, and the WER is relatively low. We also find that in Fig.\ref{fig:7a}, the speech spectrogram extracted from the entire recording is cleaner than that extracted from the blocks. It shows that even though we designed this pipeline to avoid the truncation effect caused by windowing, the pipeline that adopts blocks of mixture as input inevitably introduces noises into the extracted speech. Nevertheless, comparing the first spectrograms in Fig.~\ref{fig:7a} and Fig.~\ref{fig:7b}, we also find that if the target speech is stronger or has a higher AR value, these noises can be concealed. From the second and third spectrograms in Fig.~\ref{fig:7b}, we can hardly see any value in the latter. This happens because there is one value in the spectrogram that is large enough to conceal all the other values. After listening to the corresponding original recorded Pepper speech, we find that there is a very sharp babble noise at the end of the speech. This explains the failure to recognize words from the speech after extraction. This sharp babble noise is enhanced and shrinks the extracted speech signal during normalization. The extracted speech content becomes unrecognizable for the ASR system. In comparison, this can be alleviated by extracting target speech from blocks, because we normalize the extracted speech and concatenate it by blocks.

\subsubsection{Small Lab-based Feasibility Pilot}
As we can observe from Table~\ref{tab:2}, the proposed pipeline successfully filtered out Pepper's ego speech and restored the overlapping human speech from the overlapping mixture recorded by its microphone with a slight distortion. We also find that the computation time is less than 1 second in \textit{Thread} \RomanNumeralCaps{1}. This means that before the first audio buffer is recorded and sent to \textit{Thread} \RomanNumeralCaps{2}, the $\hat{A}_{ft}$ can be predicted and ready to be subtracted from $X_{ft}$.  The processing time for the streaming audio in \textit{Thread} \RomanNumeralCaps{2} is 10 ms, which means that the time difference between the interruption of the user and the detection of the interruption by the robot is less than 810~ms. This shows the feasibility of the proposed pipeline to extract the target speech in near-real time during HRI. Furthermore, the low mean and SD values demonstrate the reliability of the proposed pipeline. 

After further analysis, we may find a gender effect on the performance of our proposed pipeline. 5 out of 6 misinterpreted speech comes from females. This is because female participants tend to speak more softly than male participants during interruptions. This also cross-validates the analysis in Section~\ref{section:5c}. In addition, we find that incorrectly recognized words have voiceless consonants whose frequency is in the lower range of the spectrogram. This means that only restoring the information in the higher range of the spectrogram is sufficient but not enough for the ASR system. With the restored information in the higher frequency bins, a speech enhancement system should be designed to recover the values in the lower frequency bins of the spectrogram to ensure the correct interpretation of the extracted human speech.

Compared to our previous SP-based and NN-based methods, there are a few advantages of this proposed pipeline:\begin{enumerate}
    \item The proposed CNN is robust to reverberation and will not cause too much oversubtraction as the adaptive step-size method based on our test result.
    \item Compared to the end-to-end NN-based method from previous work, our method focuses more on the spectrogram estimation of the recorded robot ego speech, instead of that of the target speech. The reason is that the values in the higher frequency range that distinguish the target speech from the robot speech are too small in the spectrogram of the mixture. They may be discarded during input normalization or any normalization layers in the network. As a result, the optimization progress may just blur the entire input to achieve a local minimum loss value during training. Thus, we replace $X_{ft}$ with $R_{ft}$ as the input of the network, and it shows a great result in WER.
\end{enumerate}

\section{Conclusion and Future Work}
\label{section:6}
In this paper, we propose an ego speech filtering pipeline for robot Pepper to enable its embedded single-channel microphone to keep recording during their own speech. The proposed pipeline can filter out robot ego speech and extract human interruption speech from the recordings. We evaluated the proposed pipeline and compared its performance with our previous work \cite{c3} and two other state-of-the-art retrained TSE systems on the same dataset of overlapping robot-human speech. Our proposed pipeline exceeded all three methods. The small lab-based feasibility pilot also verified the effectiveness of our proposed pipeline in extracting human interruption speech during Pepper speech in real-time HRI. We also integrated the proposed pipeline into the SIC framework for broader accessibility to the HRI community.

Based on our analysis, there are still several limitations of the proposed pipeline's application in broader and more general scenarios. First, the values in the lower range of frequency bins are oversubtracted. This will result in incorrect translation of words that have voiceless consonants. Second, the proposed pipeline with the entire mixture as input is vulnerable to sharp babble noises. Third, the proposed pipeline with blocks as input requires the speaker to be relatively loud.

In terms of future work, we will look for possible solutions in speech enhancement to enhance the overlapping human speech extracted from blocks of mixture. Not only should this speech enhancement method be robust to sharp babble noise and be able to recover the values in the lower frequency range that have been oversubtracted by this approach, but also to process blocks as input. We also plan to do an HRI experiment on how people will prefer to interrupt a robot when they find the answers provided are wrong.

\addtolength{\textheight}{-12cm}

% \section*{ACKNOWLEDGEMENTS}
% The authors would like to thank Merle Reimann, Muhan Hou, Thomas Wiggers, and Karen Chiang for the helpful discussions.

% \addtolength{\textheight}{-12cm}


\begin{thebibliography}{99}

\bibitem{c1} Gervits, Felix, and Matthias Scheutz. "Pardon the interruption: Managing turn-taking through overlap resolution in embodied artificial agents." Proceedings of the 19th Annual SIGdial Meeting on Discourse and Dialogue. 2018.
\bibitem{c2} Aylett, Matthew Peter, and Marta Romeo. "You Don’t Need to Speak, You Need to Listen: Robot Interaction and Human-Like Turn-Taking." Proceedings of the 5th International Conference on Conversational User Interfaces. 2023.
\bibitem{c3} Yue Li, Koen V Hindriks, and Florian Kunneman. "Single-Channel Robot Ego-Speech Filtering during Human-Robot Interaction." International Symposium on Technological Advances in Human-Robot Interaction, USA, 2024.
\bibitem{c4} Skantze, Gabriel. "Turn-taking in conversational systems and human-robot interaction: a review." Computer Speech \& Language 67 (2021): 101178.
\bibitem{c5} Podpora, Michal, et al. "Human interaction smart subsystem—extending speech-based human-robot interaction systems with an implementation of external smart sensors." Sensors 20.8 (2020): 2376. 
\bibitem{c6} Nakadai, Kazuhiro, Hiroshi G. Okuno, and Hiroaki Kitano. "Humanoid active audition system improved by the cover acoustics." PRICAI 2000 Topics in Artificial Intelligence: 6th Pacific Rim International Conference on Artificial Intelligence Melbourne, Australia, August 28–September 1, 2000 Proceedings 6. Springer Berlin Heidelberg, 2000.
\bibitem{c7} Zmolikova, Katerina, et al. "Neural target speech extraction: An overview." IEEE Signal Processing Magazine 40.3 (2023): 8-29.
\bibitem{c8} Wang, Quan, et al. "Voicefilter: Targeted voice separation by speaker-conditioned spectrogram masking." arXiv preprint arXiv:1810.04826 (2018).
\bibitem{c9} Žmolíková, Kateřina, et al. "Speakerbeam: Speaker aware neural network for target speaker extraction in speech mixtures." IEEE Journal of Selected Topics in Signal Processing 13.4 (2019): 800-814.
\bibitem{c10} Subakan, Cem, et al. "Attention is all you need in speech separation." ICASSP 2021-2021 IEEE International Conference on Acoustics, Speech and Signal Processing (ICASSP). IEEE, 2021.
\bibitem{c11} Oliveira, Joao Lobato, et al. "An active audition framework for auditory-driven HRI: Application to interactive robot dancing." 2012 IEEE RO-MAN: The 21st IEEE International Symposium on Robot and Human Interactive Communication. IEEE, 2012.
\bibitem{c13} Benesty, Jacob, et al. Noise reduction in speech processing. Vol. 2. Springer Science \& Business Media, 2009.
\bibitem{c14} Upadhyay, Navneet, and Abhijit Karmakar. "Speech enhancement using spectral subtraction-type algorithms: A comparison and simulation study." Procedia Computer Science 54 (2015): 574-584.
\bibitem{c15} Wake, Naoki, et al. "Enhancing listening capability of humanoid robot by reduction of stationary ego‐noise." IEEJ Transactions on Electrical and Electronic Engineering 14.12 (2019): 1815-1822.
\bibitem{c16} Ince, Gökhan, et al. "Ego noise suppression of a robot using template subtraction." 2009 IEEE/RSJ International Conference on Intelligent Robots and Systems. IEEE, 2009.
\bibitem{c17} Jain, Sanjeev N., and Chandrashekhar Rai. "Blind source separation and ICA techniques: a review." International Journal of Engineering Science and Technology 4.4 (2012): 1490-1503.
\bibitem{c18} Hershey, John R., et al. "Deep clustering: Discriminative embeddings for segmentation and separation." 2016 IEEE international conference on acoustics, speech and signal processing (ICASSP). IEEE, 2016.
\bibitem{c19} Yu, Dong, et al. "Permutation invariant training of deep models for speaker-independent multi-talker speech separation." 2017 IEEE International Conference on Acoustics, Speech and Signal Processing (ICASSP). IEEE, 2017.
\bibitem{c20} Wang, Quan, et al. "VoiceFilter-Lite: Streaming targeted voice separation for on-device speech recognition." arXiv preprint arXiv:2009.04323 (2020).
\bibitem{c21} Ge, Meng, et al. "Spex+: A complete time domain speaker extraction network." arXiv preprint arXiv:2005.04686 (2020).
\bibitem{c22} Luo, Yi, and Nima Mesgarani. "Conv-tasnet: Surpassing ideal time–frequency magnitude masking for speech separation." IEEE/ACM transactions on audio, speech, and language processing 27.8 (2019): 1256-1266.
\bibitem{c23} Maciejewski, Matthew, et al. "WHAMR!: Noisy and reverberant single-channel speech separation." ICASSP 2020-2020 IEEE International Conference on Acoustics, Speech and Signal Processing (ICASSP). IEEE, 2020.
\bibitem{c24} Borgström, Bengt J., and Michael S. Brandstein. "Speech enhancement via attention masking network (SEAMNET): An end-to-end system for joint suppression of noise and reverberation." IEEE/ACM Transactions on Audio, Speech, and Language Processing 29 (2020): 515-526.
\bibitem{c25} Targ, Sasha, Diogo Almeida, and Kevin Lyman. "Resnet in resnet: Generalizing residual architectures." arXiv preprint arXiv:1603.08029 (2016).
\bibitem{c26} Thomas Orden, Mike EU Ligthart, Koen Hindriks, Thomas Wiggers, and Karen Chiang. [n. d.]. The Social Interaction Cloud (SIC) https://socialrobotics.atlassian.net/wiki/spaces/CBSR/overview
\bibitem{c27} Google. 2023. Google Cloud Text-to-Speech AI. https://cloud.google.com/text-to-speech
\bibitem{c28} Panayotov, Vassil, et al. "Librispeech: an asr corpus based on public domain audio books." 2015 IEEE international conference on acoustics, speech and signal processing (ICASSP). IEEE, 2015.
\bibitem{c29} S. Li, H. Liu, Y. Zhou and Z. Luo, "A SI-SDR Loss Function based Monaural Source Separation," 2020 15th IEEE International Conference on Signal Processing (ICSP), Beijing, China, 2020, pp. 356-360, doi: 10.1109/ICSP48669.2020.9321080.
\bibitem{c30} Rodrigues, Ana, et al. "Analyzing the performance of ASR systems: The effects of noise, distance to the device, age and gender." \textit{Proceedings of the XX International Conference on Human Computer Interaction}. 2019. 
\bibitem{c31} Zhivomirov, Hristo. "On the development of STFT-analysis and ISTFT-synthesis routines and their practical implementation." \textit{TEM Journal} 8.1 (2019): 56-64. 

\end{thebibliography}
\end{document}